\documentclass[11pt,twoside
]{article}

\usepackage{baaa2011}
\usepackage{graphicx}
\usepackage{subfigure}
\usepackage{psfrag}
\usepackage{amssymb}
\usepackage[spanish,activeacute,english]{babel}
\usepackage[latin1]{inputenc}
\usepackage[T1]{fontenc} % Computer Modern (CM) fonts
\usepackage{ae,aecompl} % and: dvips -Pcmz -Pamz macros_aaa.dvi
\usepackage{latexsym}
\usepackage{verbatim}
\usepackage{amsmath}
\usepackage{stmaryrd}
\usepackage{amsfonts}
\usepackage{amssymb}
\usepackage{wasysym}
\usepackage[colorlinks=true,dvips]{hyperref}
\begin{document}
%%%%%%%%%%%%%%%%%%%%%%%%%%%%%%%%%%%%%%%%%%%%%%%%%%%%%%%%%%%%%%%%%%%%%%%%%%
%%%% SELECCIONE EL IDIOMA EN QUE SE ESCRIBE EL ARTÍCULO:              %%%%
%\myselectspanish
\myselectenglish
%%%%%%%%%%%%%%%%%%%%%%%%%%%%%%%%%%%%%%%%%%%%%%%%%%%%%%%%%%%%%%%%%%%%%%%%%%
\vskip 1.0cm
\markboth{S. Torres-Flores et al. }%
{Studying the kinematics of 30 Doradus}

\pagestyle{myheadings}
%%%% DESCOMENTE LA LINEA QUE DESCRIBE EL CARACTER DE SU TRABAJO       %%%%
\vspace*{0.5cm}
%\noindent TRABAJO INVITADO 
\noindent PRESENTACIÓN ORAL
%\noindent PRESENTACIÓN MURAL
%\noindent RESUMEN 
\vskip 0.3cm
\title{A kinematic study of the giant 
star-forming region 30 Doradus}

%\title{ Template paper for publication in the Bulletin of the 
%Argentinian Astronomical Association with instructions for the use of 
%\LaTeX{}}

\author{Sergio Torres-Flores$^{1}$, 
Rodolfo Barb\'a$^{1,2}$, 
Jesús Ma\'iz Apell\'aniz$^{3}$, 
Mónica Rubio$^{4}$ 
\& Guillermo Bosch$^{5}$}

\affil{%
  (1) Departamento de F\'isica, Universidad de La Serena, Chile\\
  (2) ICATE-CONICET, San Juan, Argentina\\
  (3) Instituto de Astrof\'isica de Andaluc\'ia-CSIC, Spain\\
  (4) Departamento de Astronom\'ia, Universidad de Chile, Chile\\
  (5) Facultad de Ciencias Astron\'omicas y Geof\'isicas, UNLP, Argentina\\}

\begin{abstract} 

We present, for the first time, an optical spectroscopic data cube of the giant star-forming region 30 Doradus, obtained with the GIRAFFE on the VLT at Paranal Observatory. The main emission lines present in this data cube correspond to H$\alpha$, [NII] 6548 \rm{\AA} and 
[NII] 6584 \AA. By using this data set, we found that H$\alpha$ presents from simple to multiple profiles, which suggests that different physical mechanisms act in different ways on the excited gas in 30 Doradus. We found, at least, three unclassified large expanding structures. These structures correlate with peaks in the X-ray distribution. Given the excellent signal-to-noise ratio and the large spatial coverage of this data cube, we have studied in detail the kinematics of 30 Doradus, showing the importance of the small scale phenomena on the integrated properties of 30 Doradus

\end{abstract}

\begin{resumen}

En este trabajo presentamos, por primera vez, un cubo de datos \'optico de la region de formaci\'on estelar gigante de 30 Doradus, obtenido con GIRAFFE en el telescopio VLT. Las principales l\'ineas de emisi\'on presentes corresponden a H$\alpha$, [NII] 6548 \rm{\AA} y [NII] 6584 \AA. A partir de estos datos, encontramos que la l\'inea de H$\alpha$ presenta desde perfiles simples a componentes m\'ultiples lo que sugiere que diferentes mecanismos f\'isicos act\'uan de manera diferenciada en la cinem\'atica del gas excitado de todo el \'area. Encontramos al menos tres nuevas grandes estructuras en expansi\'on, las cuales se correlacionan con los m\'aximos de la distribuci\'on de rayos X de 30 Doradus. Debido a la excelente se\~nal-ruido y a la gran cobertura espacial de estos datos, hemos podido estudiar detalladamente la cinem\'atica de 30 Doradus, mostrando as\'i la importancia que poseen los fen\'omenos a peque\~nas escalas sobre las propiedades integradas de 30 Doradus.

\end{resumen}

\section{Introduction}

Giant H\,{\sc ii} regions (GHR) are known for containing massive young clusters, which are rich in massive stars. The strong stellar winds and the evolution of these massive stars can disrupt the interstellar medium of these GHR, resulting in the formation of super bubbles. Also, several authors have measured supersonic dispersion velocities in GHRs. The origin of these phenomena is still uncertain. The use of high resolution spectroscopic data on a near GHR is necessary to disentangle these phenomena, through the study of the kinematics of the warm gas. In this sense, one of the closest targets for doing this kind of studies is the Large Magellanic Cloud (LMC), in which lies 30 Doradus. The 30 Doradus nebula is the largest H\,{\sc ii} region in the Local Group and the most powerful source of H$\alpha$ emission in the LMC. In their core, this nebula presents a very large concentration of massive hot and luminous stars, known as R136 (see Crowther et al. 2010). The kinematics of the warm ionized gas in 30 Doradus has been analyzed using Fabry-Perot data (Laval et al. 1995) and long-slit spectroscopy by Chu \& Kennicutt (1994) and by Melnick et al. (1999), who found complex H$\alpha$ profiles. Chu \& Kennicutt (1994) found that 30 Doradus has several fast expanding shells that can not be explained using the stellar winds models. These authors suggest that SN remnants can solve that problem. Here, we present preliminary results about the kinematics of 30 Doradus, using GIRAFFE/VLT data.

\section{Observations and data reduction}

\begin{figure}[h!]
\centering
\includegraphics[width=0.40\textwidth]{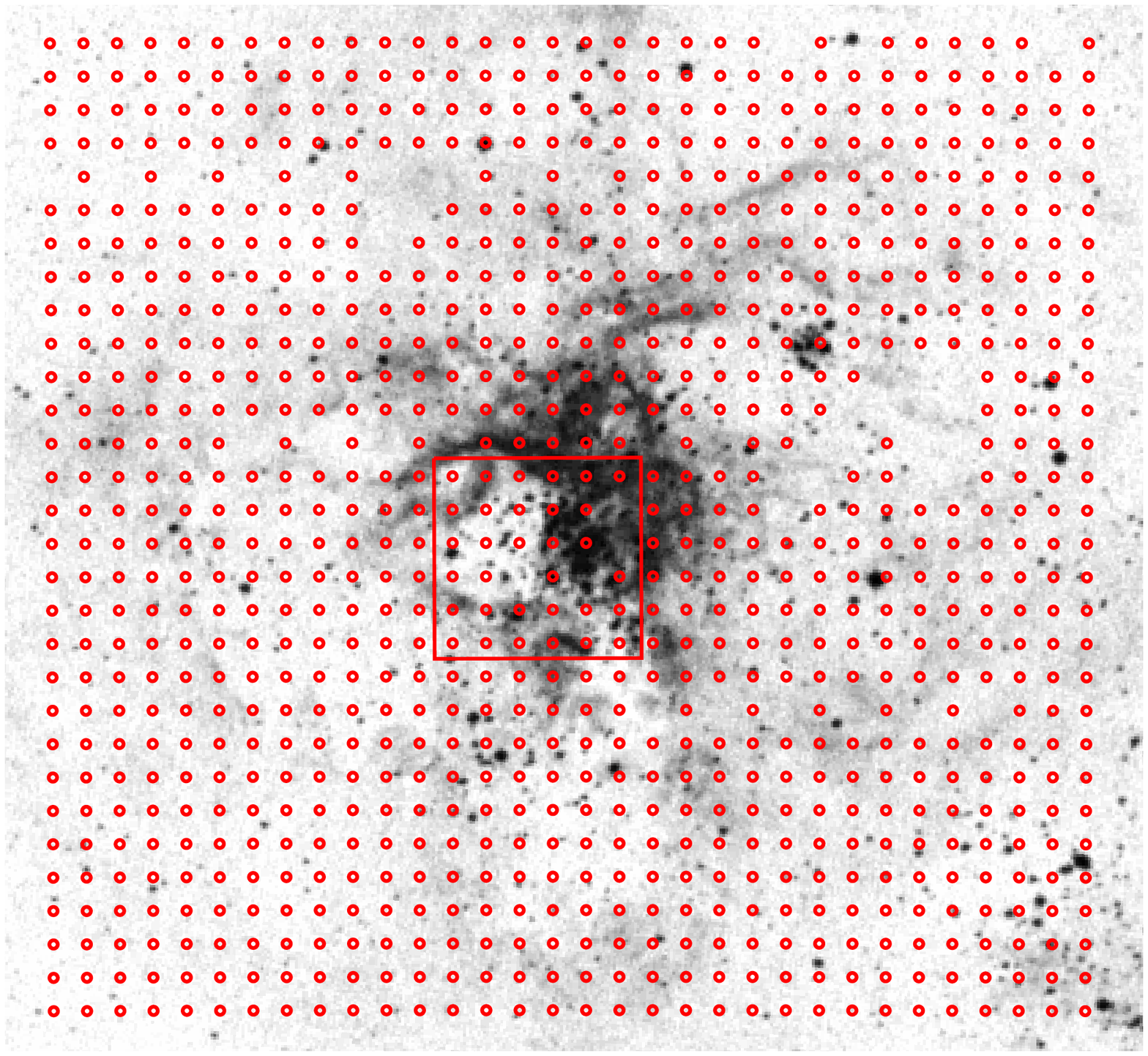}
\includegraphics[width=0.40\textwidth]{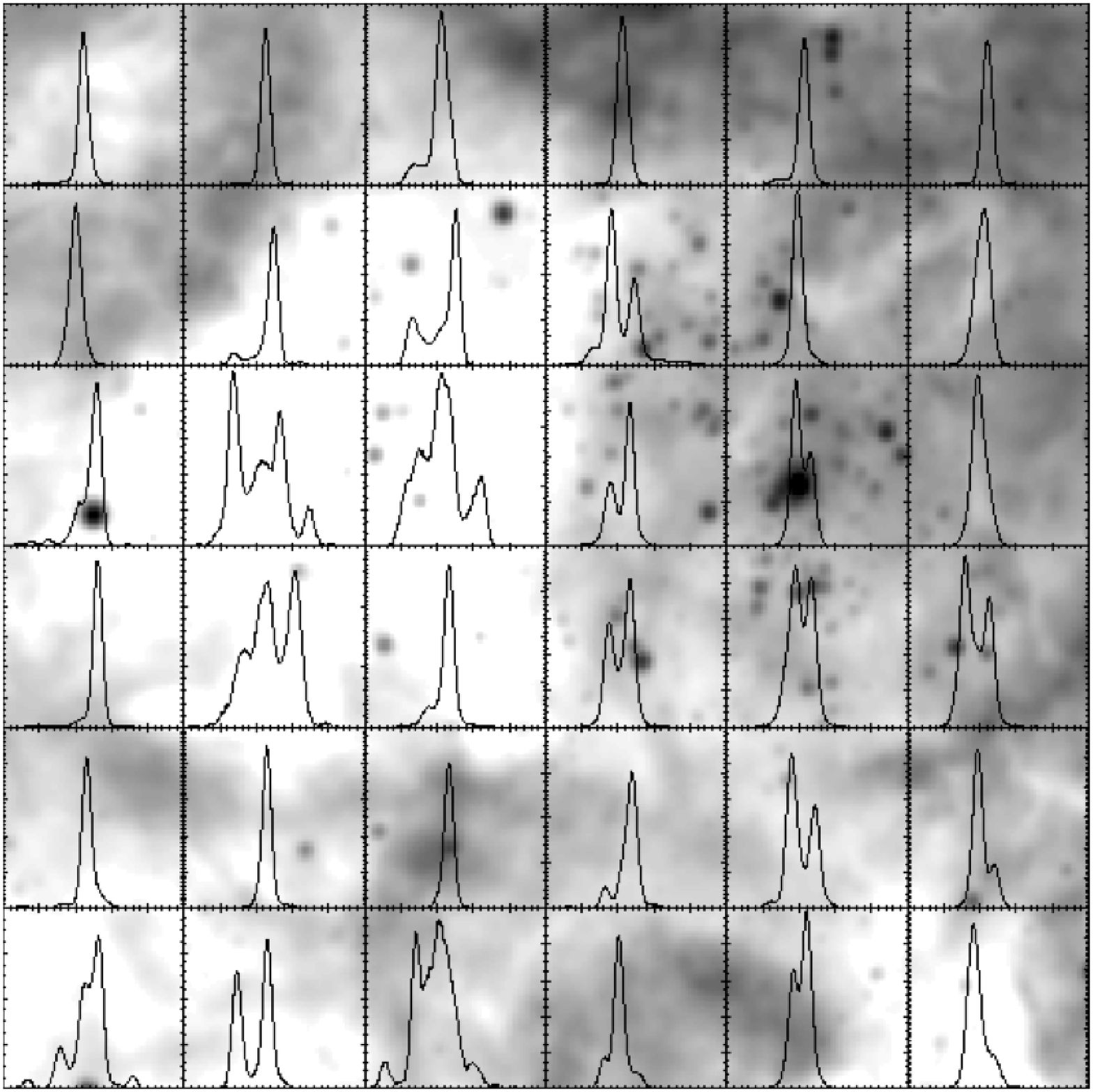}
\caption{Left panel: Red circles indicated the GIRAFFE/MEDUSA fiber positions on 30 Doradus. Right panel: H$\alpha$ profiles of the central region of 30 Doradus (indicated by the red rectangle of the left panel).}
\label{pointing}
\end{figure}

Observations of 30 Doradus were carried out using the FLAMES/GIRAFFE instrument (in the MEDUSA configuration) at the Very Large Telescope (VLT) under the high resolution mode (HR14). This allowed us to obtain about 131 spectra simultaneously, with a spectral resolution of 18 km s$^{-1}$ (FWHM). Three MEDUSA configurations were used to cover a field of view of 10'$\times$10' centered on R136, as can be seen in the left panel of Fig \ref{pointing}. At the end, we construct a grid of 32$\times$30 pixels, separated by 20 arcsec. In this case, each pixel corresponds to a MEDUSA spectrum, which allow us to obtain a spectroscopic data cube of 30 Doradus. This data cube covers from 6499 to 6691 \AA. Data reduction was performed by using the GASGANO and ESOrex softwares.

\section{Preliminary results}

\subsection{The H$\alpha$ data cube of 30 Doradus}

As noted by Chu \& Kennicutt (1994) and Melnick et al. (1999), the internal kinematics of 30 Dor are very complex. Multiple profiles are seen in most of the regions studied by these authors. In the right panel of Fig. \ref{pointing} we show the H$\alpha$ profiles of the central region of 30 Doradus, where R136 is located. From this figure, it results clear the presence of multiple profiles in 30 Doradus.

\begin{figure}[!ht]
\begin{center}
\includegraphics[width=0.45\textwidth]{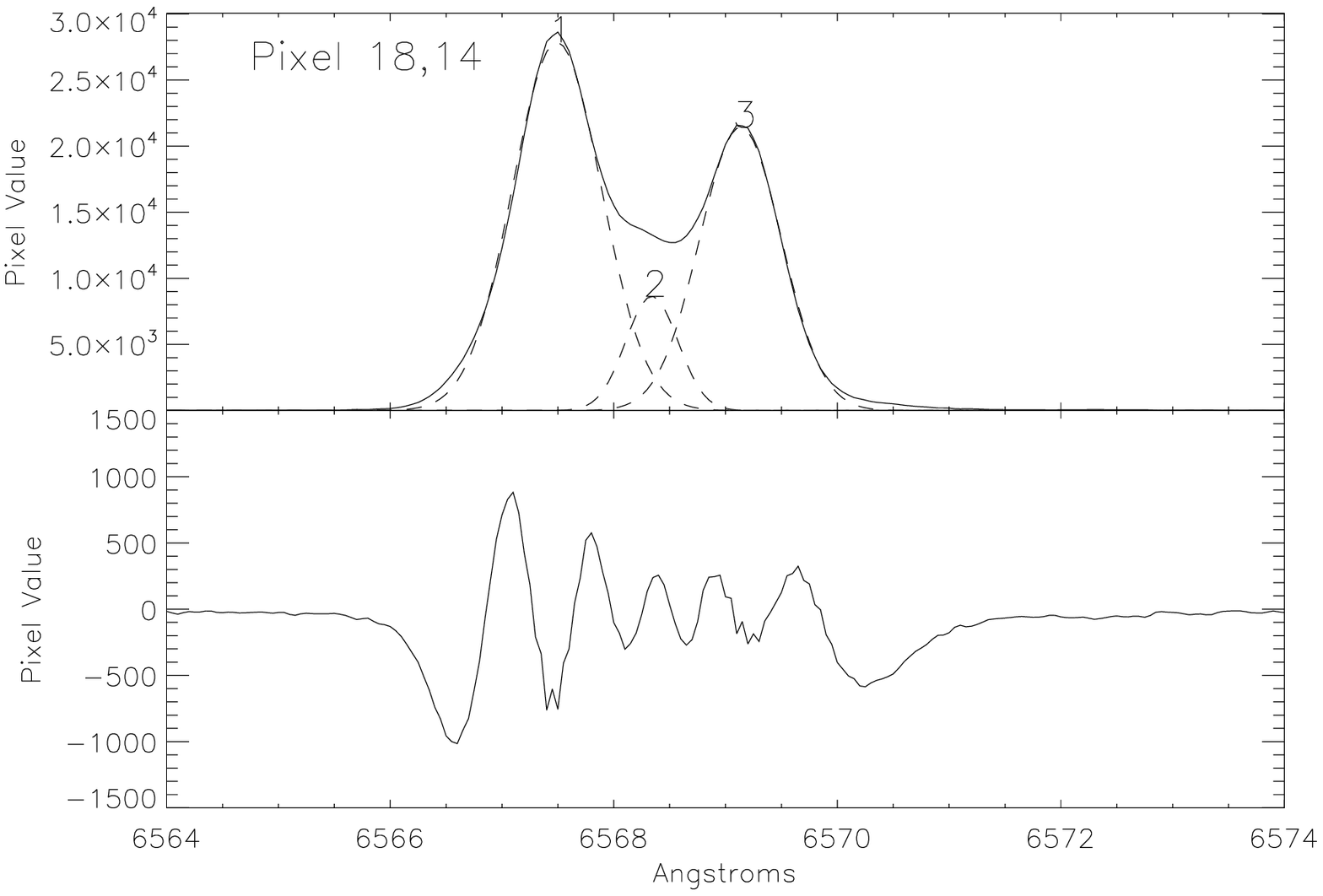}
\includegraphics[width=0.46\textwidth]{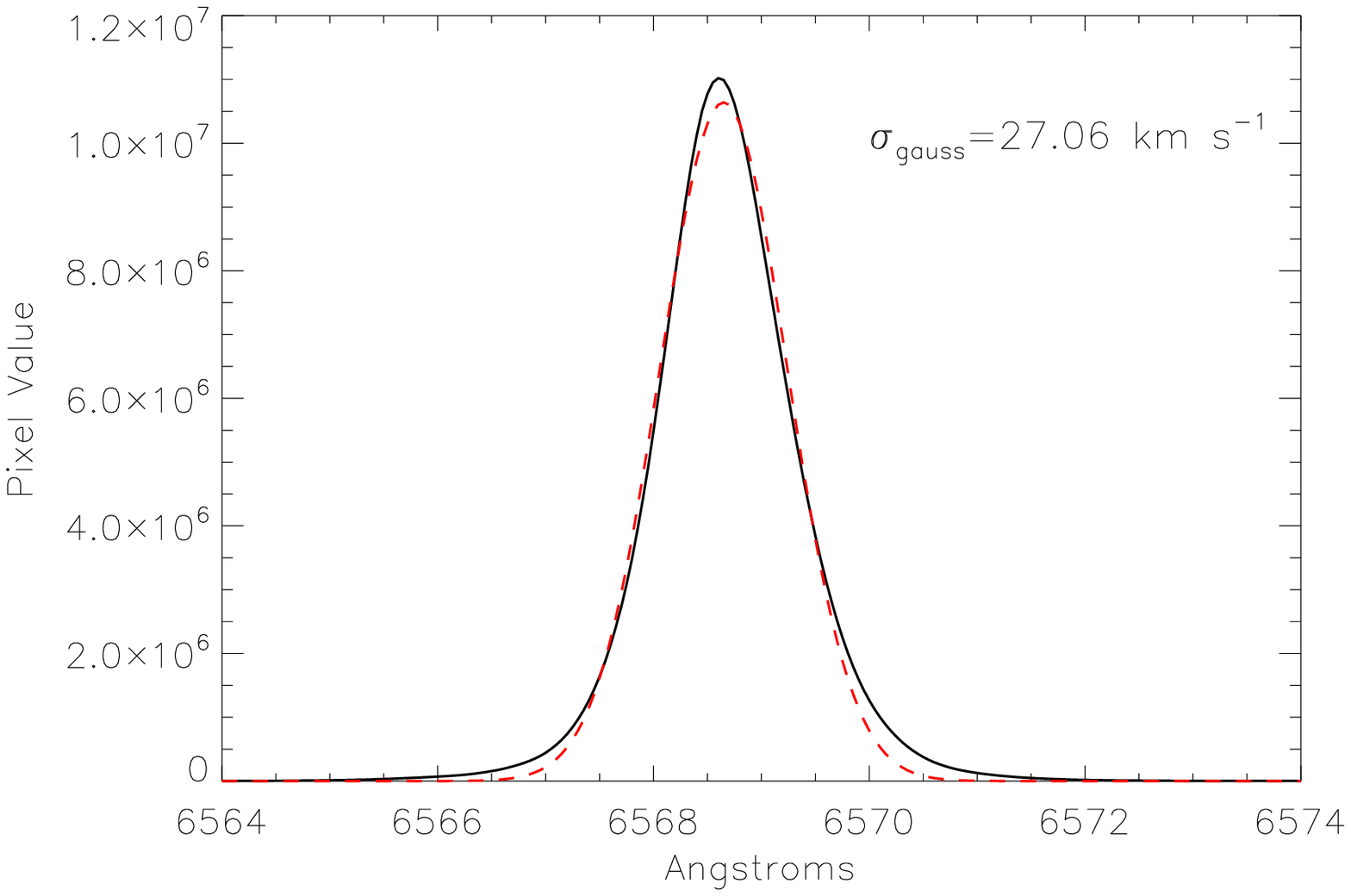}
\caption{Left panel: An example of a multiple Gaussian fitting. Right panel: Integrated H$\alpha$ profile of the data cube of 30 Doradus. The red line correspond to a Gaussian fit on the observed data.} 
\label{espectro_total}
\end{center}
\end{figure}

Given the high signal-to-noise ratio and resolution of our data we are able to study the multiple profiles by fitting Gaussians to each profile. In order to to that, we have used the routine PAN in IDL. As an example, we have applied the multiple Gaussian fitting to a region near the center of 30 Doradus (pixel 18,14). In the left panel of Fig. \ref{espectro_total} we show the result of fitting three Gaussians. We note that in this case, the residual is negligible compared with the intensity of the profiles, as can be seen in the different scales of the upper and lower left panels of Fig. \ref{espectro_total}.

In order to know the width of the integrated H$\alpha$ profile of 30 Doradus, we have fitted a single Gaussian to it. In the right panel of Fig. \ref{espectro_total} we show the integrated H$\alpha$ profile (black solid line) and the Gaussian fit to the observed data (red dashed line). The $\sigma$ of the fit is 27 km s$^{-1}$ (uncorrected for instrumental width). By inspecting the right panel of Fig. \ref{espectro_total}, it is clear the presence of wings in the observed profile. Melnick et al. (1999) suggested the presence of a low-intensity broad component to explain these wings. 

\begin{figure}[h!]
\centering
\includegraphics[width=0.35\textwidth]{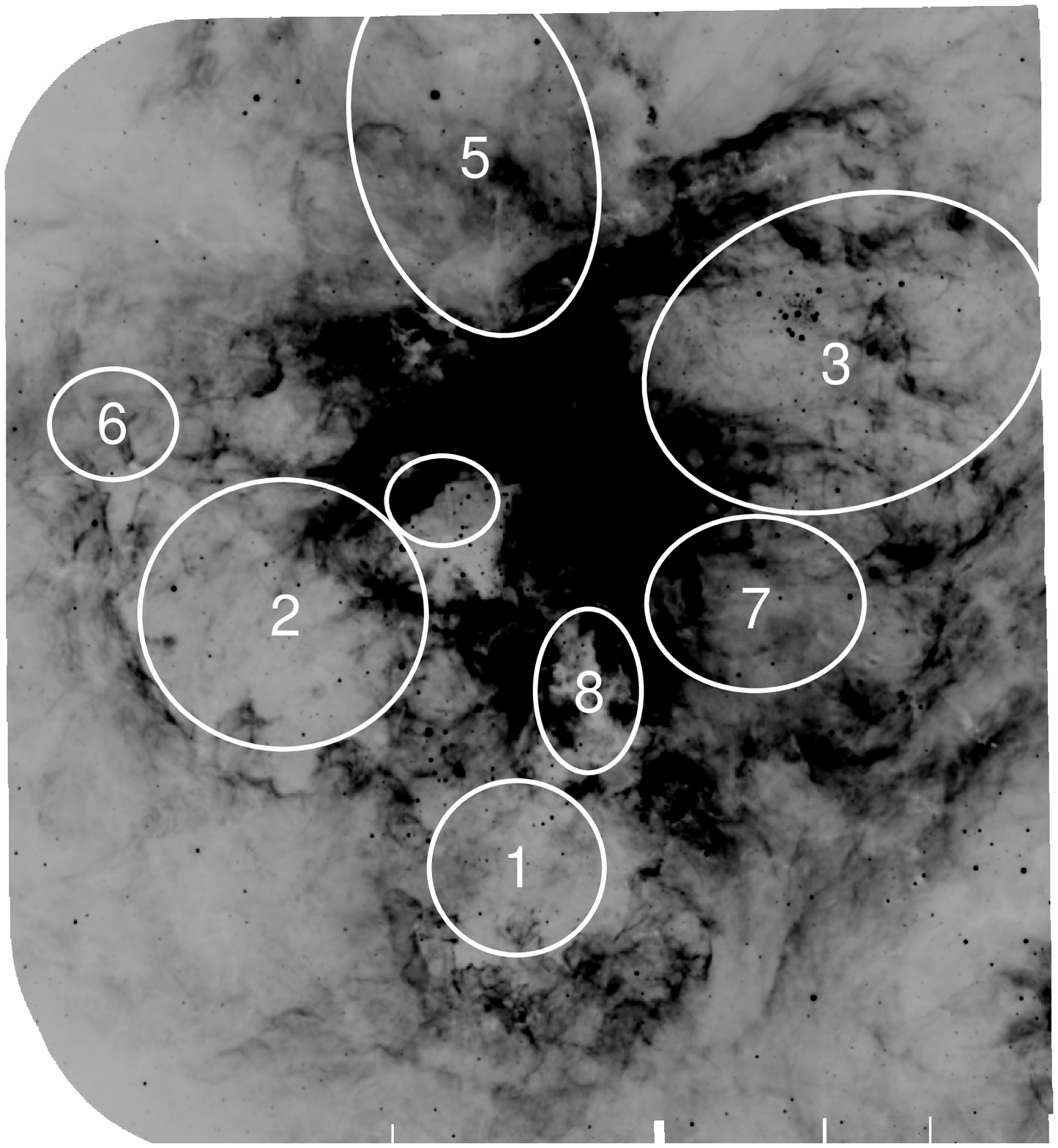}
\includegraphics[width=0.38\textwidth]{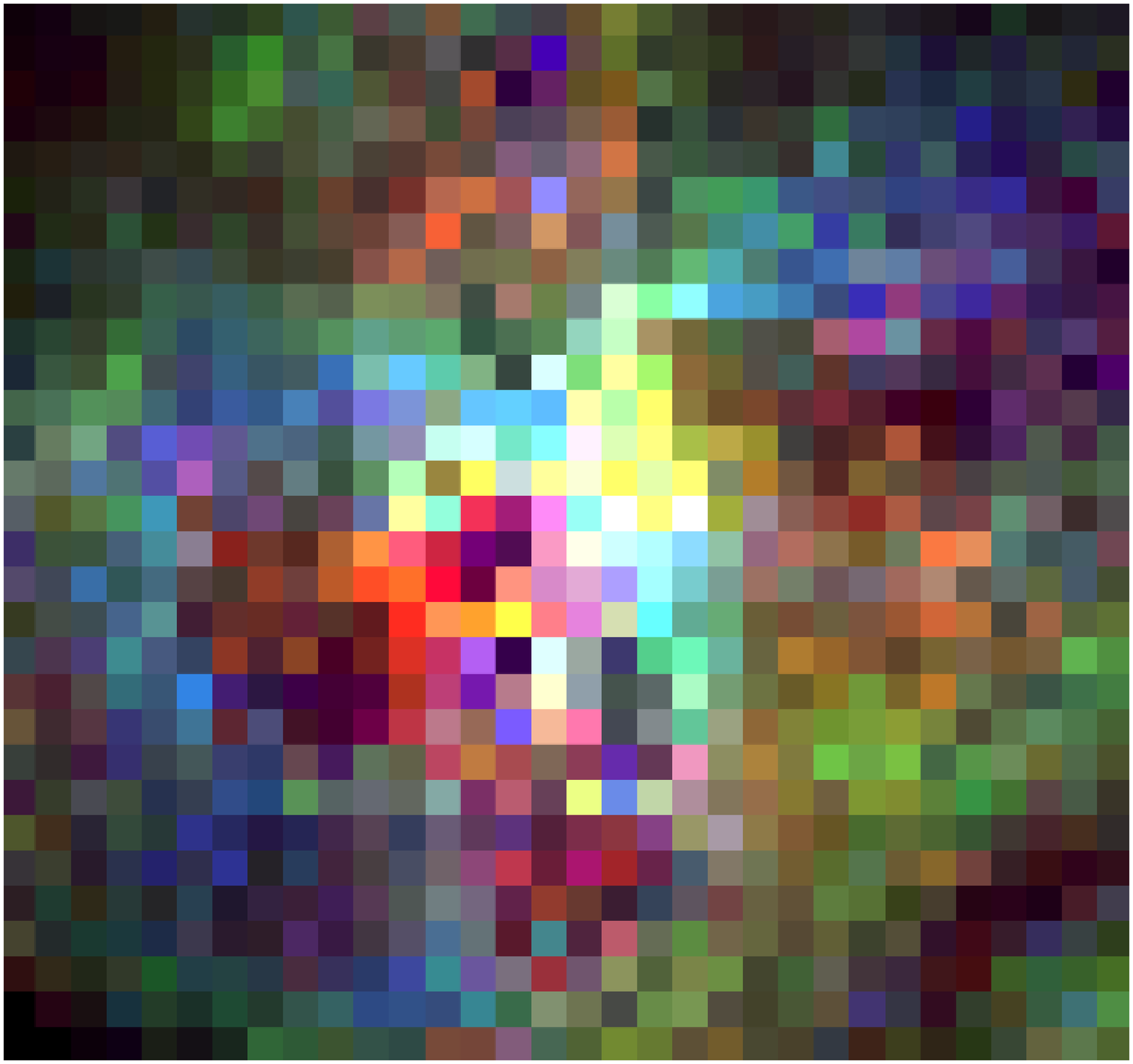}
\caption{Left panel: Large expanding structures in 30 Doradus. Right panel: Velocity map of 30 Doradus. The H$\alpha$ line was used to perform the map. In both figures the north is up and east is left.}
\label{30dor_halpha_xray_1}
\end{figure}

Chu \& Kennicutt (1994) reported the presence of several expanding structures in 30 Doradus. By inspecting the H$\alpha$ data cube of 30 Doradus, we found at least three unclassified large expanding structures. In the left panel of Fig. \ref{30dor_halpha_xray_1} we mark the expanding structures previously found by Chu \& Kennicutt (regions 1, 2, 3 and 5) and the structures found by us (regions 6, 7 and 8).

In order to show the complexity of the kinematics of 30 Doradus, we have derived a composed image of this GHR, by using the H$\alpha$ profile of each fiber position (right panel of Fig. \ref{30dor_halpha_xray_1}). Blue and red colors indicated lower and higher velocities with respect to the mean value of 30 Doradus (which has a radial velocity of 258 km s$^{-1}$ as measured from its integrated H$\alpha$ profile). In this figure, multiple profiles are indicated by pink colors.

\section{Summary}

In this work, we have presented the main features of the H$\alpha$ data cube of the local star-forming region 30 Doradus. Given the nature of this region, multiple H$\alpha$ profiles and expanding structures (linked with optical shells) are quite common. We also confirm the presence of wings in the integrated H$\alpha$ profile of 30 Doradus. These wings could be the result of a sum of several individual profiles. Given the high resolution of our data, that scenario will be tested in the near future.

\acknowledgments

We would like to thank the conference organizers for doing possible the first AAA-SOCHIAS meeting. S.T-F acknowledges the financial support of FONDECYT (Chile) through a post-doctoral position, under contract 3110087. M.R. wishes to acknowledge support from FONDECYT(Chile) grant N$^{\rm o}$ 1080335 and is supported by the Chilean {\sl Center for Astrophysics}  FONDAP N$^{\rm o}$ 15010003.
                                                                                
\begin{referencias}
\reference Chu, Y.-H., Kennicutt, R. C., Jr. 1994, \apj, 425, 720
\reference Crowther et al. 2010, MNRAS, 408, 731
\reference Laval et al. 1995, ASPC, 71, 155
\reference Melnick, J., Tenorio-Tagle, G., Terlevich, R. 1999, MNRAS, 302, 677
\end{referencias}

\end{document}